\documentclass[longauth]{aa}
\usepackage{txfonts}
\usepackage{graphicx}
\usepackage{enumerate}
\usepackage{booktabs}

\newcommand{\af}{[Mg/Fe] }

\newcommand{\vel}{km s$^{-1}$ }

\DeclareGraphicsExtensions{.ps,.pdf,.png,.jpg}

\begin{document}

\title{The Gaia-ESO Survey: Churning through the Milky Way}

\author{M. R. Hayden\inst{1}, 
A. Recio-Blanco\inst{1}, 
P. de Laverny\inst{1}, 
S. Mikolaitis\inst{2},
G. Guiglion\inst{1}, 
V. Hill\inst{1}, 
G. Gilmore\inst{3}, 
S. Randich\inst{4}, 
A. Bayo\inst{5,6}, 
T. Bensby\inst{7}, 
M. Bergemann\inst{6}, 
A. Bragaglia\inst{8}, 
A. Casey\inst{3}, 
M. Costado\inst{9}, 
S. Feltzing\inst{7}, 
E. Franciosini\inst{4}, 
A. Hourihane\inst{3}, 
P. Jofre\inst{3}, 
S. Koposov\inst{3}, 
G. Kordopatis\inst{10}, 
A. Lanzafame\inst{11,12}, 
C. Lardo\inst{13}, 
J. Lewis\inst{3}, 
K. Lind\inst{14,4}, 
L. Magrini\inst{4}, 
L. Monaco\inst{15}, 
L. Morbidelli\inst{4}, 
E. Pancino\inst{4}, 
G. Sacco\inst{4},
E. Stonkute\inst{7,2}, 
C.C. Worley \inst{3}, 
T. Zwitter\inst{16} 
}

\institute{
  Laboratoire Lagrange (UMR7293), Universit\'{e} de Nice Sophia Antipolis, CNRS, Observatoire de la C\^{o}te d'Azur, BP 4229, 06304 Nice Cedex 4, France  \label{inst1} (mhayden@oca.eu) 
  \and 
  Institute of Theoretical Physics and Astronomy, Vilnius University, Saul\.{e}tekio al. 3, LT-10257, Vilnius, Lithuania \label{inst2}
  \and 
  Institue of Astronomy, Cambridge University, Madingley Road, Cambridge CB3 0HA, UK  \label{inst3}
  \and 
  INAF-Osservatorio Astrofisico di Arcetri, Largo E. Fermi 5, 50125, Firenze, Italy  \label{inst4}
  \and 
  Instituto de F\'{i}sica y Astronom\'{i}a, Facultad de Ciencias, Universidad de Valpara\'{s}o, Av. Gran Breta\~{n}a 1111, 5030 Casilla, Valpara\'{i}so, Chile \label{inst5}
  \and 
  Max-Planck Institute for Astronomy, K\"{o}nigstuhl 17, 69117 Heidelberg, Germany \label{inst6}
  \and 
  Lund Observatory, Department of Astronomy and Theoretical Physics, Box 43, SE-221 00 Lund, Sweden \label{inst7}
  \and 
  INAF - Osservatorio Astronomico di Bologna, via Ranzani 1, 40127, Bologna, Italy \label{inst8}
  \and 
  Instituto de Astrofísica de Andaluc\'{i}a-CSIC, Apdo. 3004, 18080, Granada, Spain \label{inst9}
  \and 
  Leibniz-Institut f\"{u}r Astrophysik Potsdam (AIP), An der Sternwarte 16, 14482 Potsdam, Germany \label{inst10}
  \and 
  Dipartimento di Fisica e Astronomia, Sezione Astrofisica, Universit\'{a} di Catania, via S. Sofia 78, 95123, Catania, Italy \label{inst11}
  \and 
  INAF-Osservatorio Astrofisico di Catania, Via S. Sofia 78, 95123, Catania, Italy \label{inst12}
  \and 
  Astrophysics Research Institute, Liverpool John Moores University, 146 Brownlow Hill, Liverpool L3 5RF, United Kingdom \label{inst13}
  \and 
  Department of Physics and Astronomy, Uppsala University, Box 516, 75120, Uppsala, Sweden \label{inst14}
  \and 
  Departamento de Ciencias Fisicas, Universidad Andres Bello, Republica 220, Santiago, Chile \label{inst15}
  \and 
  Faculty of Mathematics and Physics, University of Ljubljana, Jadranska 19, 1000, Ljubljana, Slovenia \label{inst16}
}

\titlerunning{Churning Through the Milky Way}
\authorrunning{M. Hayden and Collaborators}
\abstract
{There have been conflicting results with respect to the extent that radial migration has played in the evolution of the Galaxy. Additionally, observations of the solar neighborhood have shown evidence of a merger in the past history of the Milky Way that drives enhanced radial migration. } {We attempt to determine the relative fraction of stars that have undergone significant radial migration by studying the orbital properties of metal-rich ([Fe/H]$>0.1$) stars within 2 kpc of the Sun. We also aim to investigate the kinematic properties, such as velocity dispersion and orbital parameters, of stellar populations near the sun as a function of [Mg/Fe] and [Fe/H], which could show evidence of a major merger in the past history of the Milky Way.} {We used a sample of more than 3,000 stars selected from the fourth internal data release of the Gaia-ESO Survey. We used the stellar parameters from the Gaia-ESO Survey along with proper motions from PPMXL to determine distances, kinematics, and orbital properties for these stars to analyze the chemodynamic properties of stellar populations near the Sun.}{Analyzing the kinematics of the most metal-rich stars ([Fe/H]$>0.1$), we find that more than half have small eccentricities ($e<0.2$) or are on nearly circular orbits. Slightly more than 20\% of the metal-rich stars have perigalacticons $R_p>7$ kpc. We find that the highest [Mg/Fe], metal-poor populations have lower vertical and radial velocity dispersions compared to lower [Mg/Fe] populations of similar metallicity by $\sim10$ km s$^{-1}$. The median eccentricity increases linearly with [Mg/Fe] across all metallicities, while the perigalacticon decreases with increasing [Mg/Fe] for all metallicities. Finally, the most [Mg/Fe]-rich stars are found to have significant asymmetric drift and rotate more than 40 \vel slower than stars with lower [Mg/Fe] ratios.}{While our results cannot constrain how far stars have migrated, we propose that migration processes are likely to have played an important role in the evolution of the Milky Way, with metal-rich stars migrating from the inner disk toward to solar neighborhood and past mergers potentially driving enhanced migration of older stellar populations in the disk.}

\maketitle
\section{Introduction} 
With the ability to resolve individual stars and stellar populations, the Milky Way is an ideal test bed for models of chemical and galaxy evolution. In particular, the chemical and kinematic properties of different stellar populations can hold key insights into the role past mergers or secular processes such as migration have had on the evolution of the Milky Way and galaxies in general. Large-scale systematic surveys of the Milky Way from several groups,    such as SEGUE \citep{Yanny2009}, RAVE \citep{Steinmetz2006}, APOGEE \citep{Majewski2015}, HERMES-GALAH \citep{Freeman2010}, and Gaia-ESO \citep{Gilmore2012} , are currently underway to determine the stellar structure of the Galaxy. The relative importance of various secular processes such as blurring and radial migration in galaxy evolution is a crucial matter of debate (e.g., \citealt{Sellwood2002,Schonrich2009,Haywood2016}). To disentangle the proposed scenarios, the study of Galactic disk stellar populations can provide important constraints through the analysis of chemodynamical correlations.

The study of stellar populations within the Milky Way has led to key insights into its structure. The thick disk was discovered as an overdensity of stars in the solar neighborhood at large heights above the plane \citep{Gilmore1983}. The thick disk has been observed to be enhanced in $\alpha$ elements and is distinct in the [$\alpha$/Fe] versus [Fe/H] plane (e.g., \citealt{Fuhrmann1998,Bensby2003,Reddy2006,Lee2011,Recio-Blanco2014}). A relation between the $\alpha$ abundance and the age of a star has been observed in $\alpha$-enhanced populations in the solar neighborhood \citep{Haywood2013}, where older stars are significantly $\alpha$-enhanced compared to younger stars. The thick disk was initially thought of as only metal poor, but more recent observations have shown that this disk extends to much higher metallicities, potentially up to super-solar abundances (e.g., \citealt{Bensby2007,Adibekyan2012}), although the separation between the thin and thick disk is not entirely clear (e.g., \citealt{Adibekyan2012,Bensby2014,Hayden2015}). The eccentricities of the thick disk populations are highest for more metal-poor thick disk stars, and the median eccentricity gradually decreases as metallicity increases \citep{Recio-Blanco2014}.

The structure of the disk has evolved with time, as secular processes such as radial mixing have caused stars to stray significantly from their birth radii.  There are two main mechanisms that contribute to radial mixing: blurring and migration (churning). Blurring is the epicyclic motion of stars from their birth radii because they are on eccentric orbits, but the angular momentum of their original orbit is unchanged with time. Radial migration is when the angular momentum of a star is changed due to interactions with transient spiral arms or the Galactic bar, causing the star to move to smaller or larger galactic radii depending on the change in momentum. This mechanism can transport stars from the inner and outer disk to the solar neighborhood while preserving the eccentricity of the orbit. The relative magnitude of blurring and migration on the evolution of the Milky Way is not currently well understood, with many contradictory results (see, e.g., \citealt{Sellwood2002,Schonrich2009,Loebman2011,Minchev2013,Haywood2013,Haywood2016,Loebman2016}. The presence of extremely metal-rich stars ([Fe/H]$>0.1$) in the solar neighborhood is a key signature of radial mixing. The interstellar medium (ISM) of the solar neighborhood is $\sim$solar, so metal-rich stars observed in the solar neighborhood likely did not form here. These populations must have arrived through blurring or migration. \citet{Kordopatis2015} find that the bulk of the super-solar metallicity stars in the solar neighborhood have eccentricities less than 0.15, or are on roughly circular orbits. This makes migration an attractive option for how these stellar populations arrived in the solar neighborhood, as the difference between peri- and apogalacticon is not dramatic for these orbits.

Mergers also have the ability to reshape the disk or enhance radial mixing (e.g., \citealt{Bird2012,Widrow2012,Minchev2014}). Mergers or satellite flybys can induce bending waves in the disk \citep{Widrow2012}. These bending waves can drive blurring of stellar populations \citep{Bird2012}. However, mergers and satellite interactions can also drive or enhance spiral structures in the disk of the primary halo, causing increased radial migration \citep{Minchev2014}. The impact of mergers tends to be largest in the outer disks of galaxies, as the potential is lower than in the inner parts of the disk. However, the final state of the structure of the disk after a merger depends on the mass ratio, gas fraction, velocity, and inclination (e.g., \citealt{Hopkins2008,Donghia2016}). It is possible that some kinematic signatures of the past merger history of the Milky Way can be observed in the disk today.
 
Recent studies of the thick disk in the solar neighborhood have revealed clues about the merger history of the Milky Way \citep{Minchev2014}. Using observations taken from RAVE, \citet{Minchev2014} have found that the velocity dispersion is lower for the most $\alpha$-enhanced (oldest thick disk stars; see \citealt{Haywood2013}) populations, which runs counter to the idea that secular processes increase the velocity dispersion of a stellar population over time. These authors have claimed that these observations are consistent with a merger early on in the history of the Milky Way and that this merger leads to enhanced migration, which predominately affects the kinematically coolest stellar populations. A merger in a chemodynamic simulation from \citet{Minchev2013} also shows the same signatures on the observed velocity dispersion in the solar neighborhood. This merger has a 1:5 mass ratio and occurs at z$\sim2$ with low inclination and causes spiral structures to form in the disk. These merger-enduced spiral structures drive enhanced migration throughout the disk. This result was also observed in \citet{Guiglion2015} with observations taken from the second internal data release of the Gaia-ESO survey, although the signal was less obvious in the Gaia-ESO observations compared to RAVE.

We use observations taken from the fourth internal data release (iDR4) of the Gaia-ESO survey to shed new light on the potential impact of radial migration and mergers on the evolution of the Milky Way through characterizing the chemodynamic structure of the nearby disk. The Gaia-ESO spectroscopic survey provides precision chemical abundances for a large sample of stars throughout the Milky Way disk, which when combined with proper motions from PPMXL \citep{Roeser2010} provide one of the best pictures of the chemodynamic structure of the nearby disk. This paper is organized as follows. In Section 2 we describe the Gaia-ESO data set, in Section 3 we present our results on the velocity dispersion and orbital characteristics for this sample, and in Section 4 we discuss the implications of our results on the evolutionary history of the Milky Way.

\section{Data and sample selection}

Data are taken from the fourth internal data release (iDR4) of the Gaia-ESO survey \citep{Gilmore2012}. We select main sequence stars ($\log{g}>3.8$) that are observed with the HR10 ($5339-5619 \text{\AA}, R\sim19800$) and H21 ($8484-9001 \text{\AA}, R\sim16200$) GIRAFFE setups with reliable stellar parameters, radial velocities, and PPMXL proper motions \citep{Roeser2010}. Stellar parameters are determined using an average between several different methods, i.e., SME \citep{Piskunov2016}, Ferre \citep{AllendePrieto2005}, and Matisse \citep{Recio-Blanco2006}. Stars are required to have a minimum signal-to-noise ratio of 20, along with errors in stellar parameters less than $\sigma_{\textrm{T}_\textrm{eff}}<200$K, $\sigma_{\log{g}}<0.4$ dex, $\sigma_{\textrm{[Fe/H]}}<0.15$ dex, $\sigma_{\textrm{[Mg/Fe]}}<0.1$ dex. The quoted errors in the stellar parameters are the dispersion of the measurements of the methods mentioned above. For our sample of 3,998 stars, we use Mg as our measurement for the $\alpha$ abundance of a star, as the separation between high- and low-$\alpha$ populations is most obvious with Mg compared to the other $\alpha$ elements.



\begin{figure}
\centering
\includegraphics[width=3.0in]{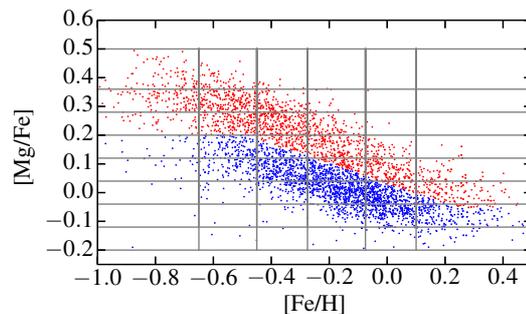}
\caption{[Mg/Fe] versus [Fe/H] for the subset of DR4 used in this paper. The gray lines denote the different metallicity and [Mg/Fe] bins used in later sections. We separate populations likely belonging to the thin and thick disks using the [Mg/Fe] ratio, with higher [Mg/Fe] stars likely belonging to the thick disk denoted in red and lower [Mg/Fe] stars likely belonging to the thin disk in blue.}
\label{survey}
\end{figure}

\subsection{Distances}
Distances were computed for the entire iDR4 sample with methods similar to those employed by \citet{Kordopatis2011} and \citet{Recio-Blanco2014}. In this method, the distance was computed by comparing the observed stellar parameters (T$_{\textrm{eff}}$, $\log{g}$, [Fe/H], and apparent magnitude) to PARSEC isochrones \citep{Bressan2012}. The isochrones span a range of metallicities from -2.5 to +0.6 in steps of 0.1 dex in [Fe/H] and ages from $\log{\tau}=8$ to $\log{\tau}=10.10$ in steps of $\tau=0.05$ dex. The isochrones use a Chabrier IMF \citep{Chabrier2003} and we assume a constant star formation history to determine the luminosity function. Isochrone grid points within $2\sigma$ of the observational errors are used and weighted according to the errors in the individual stellar parameters and luminosity function. Errors in the distances were computed using 1000 Monte Carlo (MC) runs using the observational uncertainties for each star. 
\subsection{Velocities and orbits}

Space velocities were computed using the equations as described in the Appendix of \citet{Williams2013}. We assumed a solar motion relative to the local standard of rest of (8.5, 13.38, 6.49) (U$_\odot$,V$_\odot$,W$_\odot$, \citealt{Coskunoglu2011}) and a circular velocity of $V_c=220$ km s$^{-1}$. Errors in the velocity were determined using 1,000 MC runs of the distance and proper motion errors. For each velocity component, we keep only stars with errors $\sigma_{v_i}<25$ km s$^{-1}$, so there are different numbers of stars in each velocity subsample as shown in Table \ref{tabsample}. We have roughly double the number of stars in each subsample compared to the analysis by \citet{Guiglion2015} in iDR2. The velocity dispersion for each subsample is measured using the maximum likelihood method described in \citet{Guiglion2015}. 
\begin{equation}
\small{L(\mu,\sigma_i) = \prod_{i=1}^{N}\frac{1}{\sqrt{2\pi(\sigma_{i}^{2}+ev_{i}^{2})}}\exp\left({-\frac{1}{2}\frac{(v_i-\mu)^2}{2(\sigma_i^2+ev_i^2)}}\right)}
,\end{equation}
where $\sigma_i$ is the velocity dispersion, $\mu$ is the average velocity, $v_i$ is the velocity of a given star, and $ev_i$ is the error in the velocity. We minimize the log-likelihood $\Lambda\equiv2\ln{L}$ by solving, where the partial derivatives of $\Lambda$ are zero, i.e., $\frac{\delta\Lambda}{\delta\mu}$ and $\frac{\delta\Lambda}{\delta\sigma_i}$, which gives
\begin{equation}
  \small{\sum_{i=1}^{N} \frac{v_i}{(\sigma_i^2+ev_i^2)}-\mu\sum_{i=1}^{N}\frac{1}{\sigma_i^2+ev_i^2}=0}
\end{equation}
\begin{equation}
  \small{\sum_{i=1}^{N} \frac{(\sigma_i^2+ev_i^2)^2-(v_i-\mu)^2}{(\sigma_i^2+ev_i^2)^2}=0}
;\end{equation}

see also \citet{Godwin1987,Pryor1993}.
\par
We computed orbits for a smaller subset of 2,364 stars, which had reliable measurements ($\sigma_i<30 \textrm{km s}^{-1}$) in all velocity components. Because reliable measurements are needed in all the velocity components, we slightly relaxed the uncertainties in the velocity to 30 \vel for the orbital measurements to ensure a robust sample size. The orbits were computed with \textit{Galpy} \citep{Bovy2015a} using a Milky Way potential consisting of a bulge, disk, and NFW halo (\citealt{Navarro1997}; the default MWPotential2014 from \textit{Galpy}, see \citealt{Bovy2015a} for details). Errors in the orbital parameters were determined using 1000 MC runs. 

\begin{table}
\label{tabsample}
\caption{Number of stars in iDR4 in each stellar subsample.}
\centering
\begin{tabular}{c c}
\hline
\hline
Stellar sample & $\textrm{N}_{\textrm{Stars}}$ \\
\hline
$\sigma_{V_z}$ & 2632 \\
$\sigma_{V_r}$ & 2613 \\
$\sigma_{V_\phi}$ & 2851 \\
$\textrm{[Fe/H]}>0.1$ & 198 \\
$\textrm{[Fe/H]}>0.1$,$R_p>7$ kpc & 51 \\
$\textrm{[Fe/H]}>0.25$ & 36 \\
$\textrm{[Fe/H]}>0.25$,$R_p>7$ kpc & 7 \\
\hline 
\end{tabular}
\end{table}

\section{Results}
\subsection{Velocity distributions}

\begin{figure*}[t!]
\centering
\includegraphics[width=6.3in]{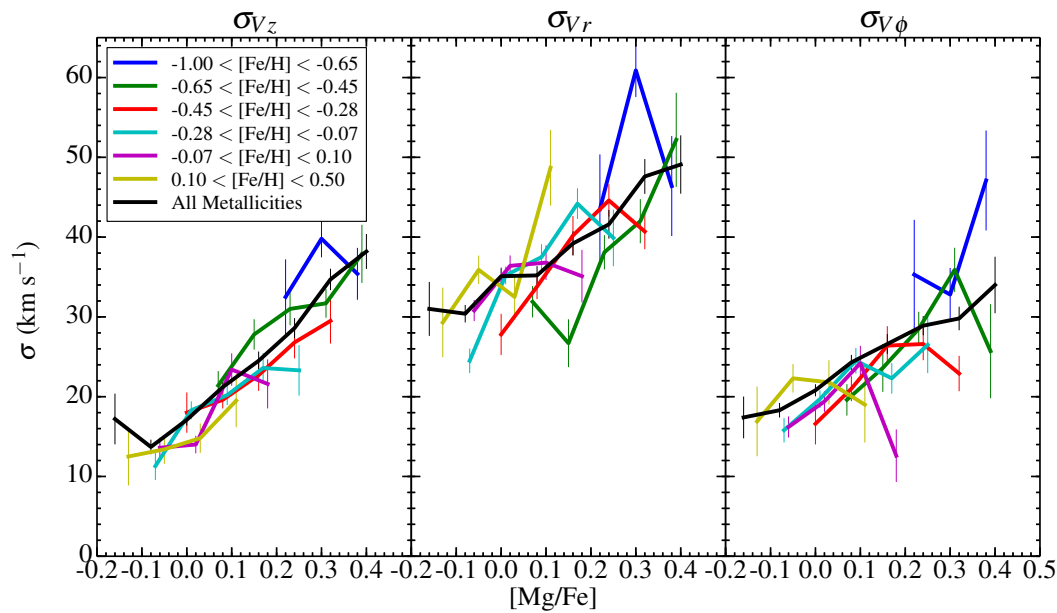}
\caption{Velocity dispersion as a function of \af for different metallicities. The velocity dispersions are calculated with bins of 0.08 dex in [$\alpha$/Fe], but are shown slightly offset above. Globally (black lines), the velocity dispersion increases with \af. However, for many individual metallicity bins the highest-$\alpha$ populations show a decrease in velocity dispersion relative to the lower-$\alpha$ populations of the same metallicity.}
\label{vdisp}
\end{figure*} 

We measured the velocity dispersion as a function of [Mg/Fe] abundance and metallicity in bins of 0.08 dex in [Mg/Fe] and a range of metallicities between $-1.0<\textrm{[Fe/H]}<0.5$ as noted in Fig. \ref{survey}. We used only those [Mg/Fe] and [Fe/H] bins with at least 20 stars and removed the largest velocity outliers from the sample (more than $2\sigma$ from the median velocity). Errors in the velocity dispersion measurements were determined via bootstrapping. 

We find that the vertical velocity dispersion $V_z$ increases linearly with increasing [Mg/Fe]-enhancement for most of the stellar populations in our sample and indeed for the bulk sample, as shown in Fig. \ref{vdisp}. The most metal-poor component (blue line) has a steep increase in velocity dispersion $0.25<$\af$<0.35$ of $\sim10$ \vel. However, there is a dip of more than 6 \vel for the highest [Mg/Fe] metal-poor populations. For stars with metallicities between $-0.65<\textrm{[Fe/H]}<-0.4$ we find a roughly linear increase in vertical velocity dispersion with \af. However, if we allow for a less stringent error criteria, $\sigma_{v_i}<30$ \vel, we find a somewhat peculiar increase in vertical velocity dispersion for these stars with a large jump of $\sim12$ \vel at intermediate \af values. This feature is not present with the more stringent cut of 25 \vel on the precision of the velocity measurements shown in Fig. \ref{vdisp}.

The trends in radial velocity dispersion mirrors that of the vertical velocity dispersion with the bulk sample increasing linearly with [Mg/Fe]. The velocity dispersion decreases for the most metal-poor, [Mg/Fe]-enhanced stars by $\sim14$ km s$^{-1}$, which is similar to the decrease observed in the vertical velocity dispersion. There is also a large spike in the most [Mg/Fe]-rich metal-rich stars in radial velocity dispersion of $\sim20$ km s$^{-1}$. 

The rotational velocity dispersion shows a linear trend for the bulk sample similar to that of the vertical component with the velocity dispersion increasing with [Mg/Fe] enhancement. However, for the highest [Mg/Fe] metal-poor populations we do not see the inversion in rotational velocity dispersion that was observed in the vertical component, but instead see a jump of $\sim12$ \vel for these stars.

Additionally, there is a general trend of having a turnover for many of the most [Mg/Fe]-rich bins at all metallicities in the different components or at least a reversal of the increasing velocity dispersion trend with [Mg/Fe]. The trends in all components are similar to those observed with earlier Gaia-ESO observations from \citet{Guiglion2015} and the global trends found there are observed in the newest data set.

\begin{figure}[t!]
\centering
\includegraphics[width=3in]{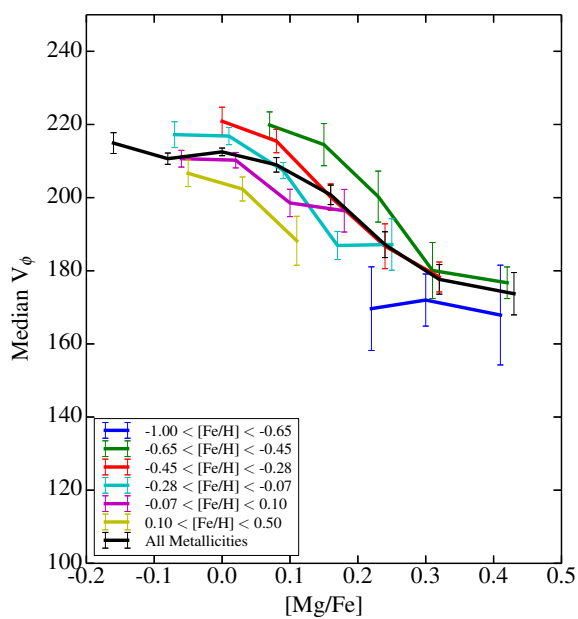}
\caption{Median V$_{\phi}$ as a function of [Mg/Fe] and [Fe/H]. For the most metal-poor populations, we see a clear separation between the lowest [Mg/Fe] populations that have roughly thin disk kinematics, while the higher [Mg/Fe] populations have significantly lower rotational velocities.}
\label{medvel}
\end{figure}

We examined the median velocities of the different components to see if there are any significant trends present in the iDR4 dataset. Errors in the median velocities were determined with bootstrapping. We find that the median velocity of the radial and tangential components scatter around zero, regardless of [Fe/H] and [Mg/Fe]. However, we find clear differences in the rotational velocity component at different metallicities and [Mg/Fe] abundances as shown in Fig. \ref{medvel}, which has been observed previously (e.g., \citealt{Lee2011,Recio-Blanco2014,Guiglion2015}). There is significant asymmetric drift observed in our populations, in which the higher-[Mg/Fe] populations have lower rotational velocities than the lower-[Mg/Fe] populations of the same metallicity. In the case of the relative metal-poor stars (green and red curves), the highest-[Mg/Fe] populations have rotational velocities 40 \vel lower than the lowest-[Mg/Fe] stars. The trend of decreasing rotational velocities with [Mg/Fe] appears to level out for the very highest \af bin. 

\subsection{Orbital characteristics}

\begin{figure}[t!]
\centering
\includegraphics[width=3in]{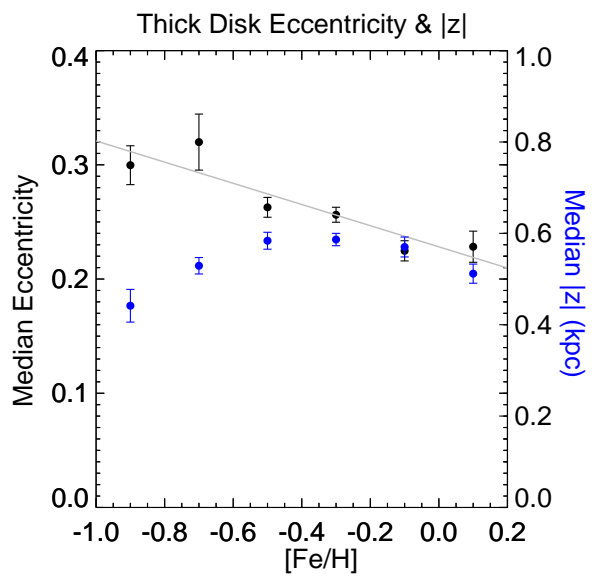}
\caption{Median eccentricity as a function of [Fe/H] for thick disk stellar populations denoted in Fig. \ref{survey}. There is a general trend of decreasing median eccentricity with [Fe/H]. }
\label{thickecc}
\end{figure}

As described in Section 2.2, we calculated different orbital properties for a smaller subset of stars with precise measurements in all three kinematic components. The eccentricity distribution for the thick disk has been previously observed to have a strong dependence on metallicity \citep{Recio-Blanco2014}. We measured the median eccentricity of the thick disk (red stars in Fig. \ref{survey}) as a function of metallicity, as shown in Fig. \ref{thickecc}. For these high-[Mg/Fe] populations, eccentricity decreases fairly consistently with metallicity, starting at $\sim0.35$ for the most metal-poor thick disk stars and gradually decreasing to $\sim0.2$ for the more metal-rich thick disk populations. To check for possible observational biases, such as sampling stars at different distances from the midplane, we also plot the median height above the plane for the thick disk stars. We find that it does not change appreciably with metallicity. The median height above the plane is between 400 and 600 pc across the metallicity space shown in Fig. \ref{thickecc}, so the eccentricity trend in the thick disk populations is not due to dramatic changes in distance from the plane for our sample as a function of [Fe/H].

\begin{figure}[t!]
\centering
\includegraphics[width=3in]{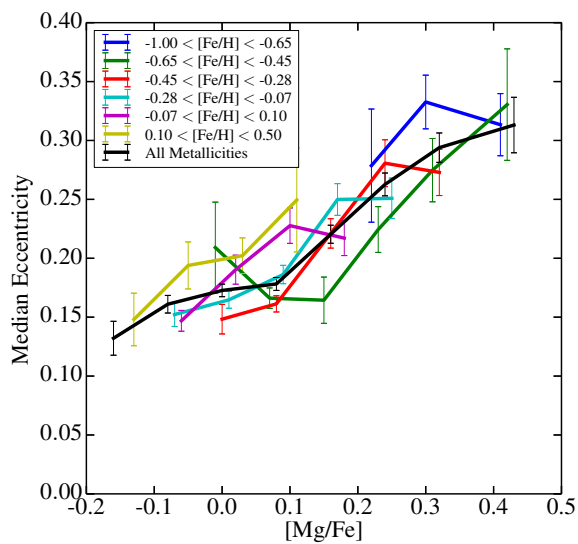}
\caption{Distributions of eccentricity for stars between a range of metallicity $-1.0<\textrm{[Fe/H]}<0.5$ as a function of [Mg/Fe]. The lower [Mg/Fe] stars have a thin disk eccentricity distribution with the bulk of the stars on roughly circular orbits. The higher [Mg/Fe] stars have a more thick disk-like distribution with the bulk of the stars having non-circular orbits.}
\label{orbitpoor}
\end{figure}

We also examine in Fig. \ref{orbitpoor} how eccentricity changes as a function of [Mg/Fe] for different metallicities. We find that the median eccentricity for low-[Mg/Fe] populations is generally low, with $e\sim0.15$, although it is slightly higher in metal-rich solar-[Mg/Fe] populations. However, we see a clear global trend that the median eccentricity increases as \af increases for all metallicities. Of particular interest is the eccentricity distribution for the highest \af populations in each metallicity bin to further test the hypothesis of a past merger driving enhanced migration. We find that for many of the metallicity bins, the median eccentricity decreases for the highest \af stars of the same metallicity, although this is within the errors of our measurements. 

\begin{figure}[t!]
\centering
\includegraphics[width=3in]{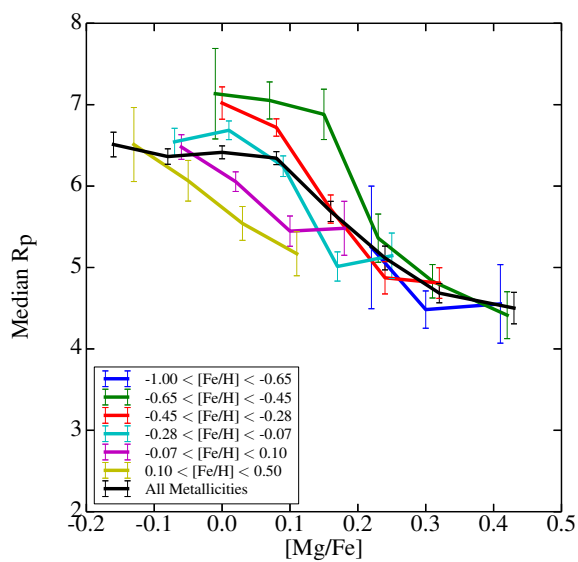}
\caption{Distributions of perigalacticon across a range of metallicity between $-1.0<\textrm{[Fe/H]}<0.5$ as a function of [Mg/Fe]. The lower [Mg/Fe] stars have R$_P$ concentrated in the solar neighborhood, while the bulk of the higher [Mg/Fe] stars have orbits that take them to the inner disk.}
\label{periall}
\end{figure}

We finally examined the distribution of perigalacticons for these same metal-poor populations in Fig. \ref{periall}. The lowest [Mg/Fe] stars have perigalacticons peaked in the solar neighborhood, matching their roughly circular orbital distribution closely. This is in stark contrast to the higher [Mg/Fe] stars, with                           many stars reaching the inner disk and having median perigalacticons $R_p<5$ kpc. We also observe what appears to be a discontinuity between the thin and thick disk in the perigalacticon distribution, with a large decrease in perigalacticon in the transition between thin and thick disk populations ($0.1<\alpha<0.2$). 

\subsection{Radial migration}

\begin{figure*}[t!]
\centering
\includegraphics[width=6.3in]{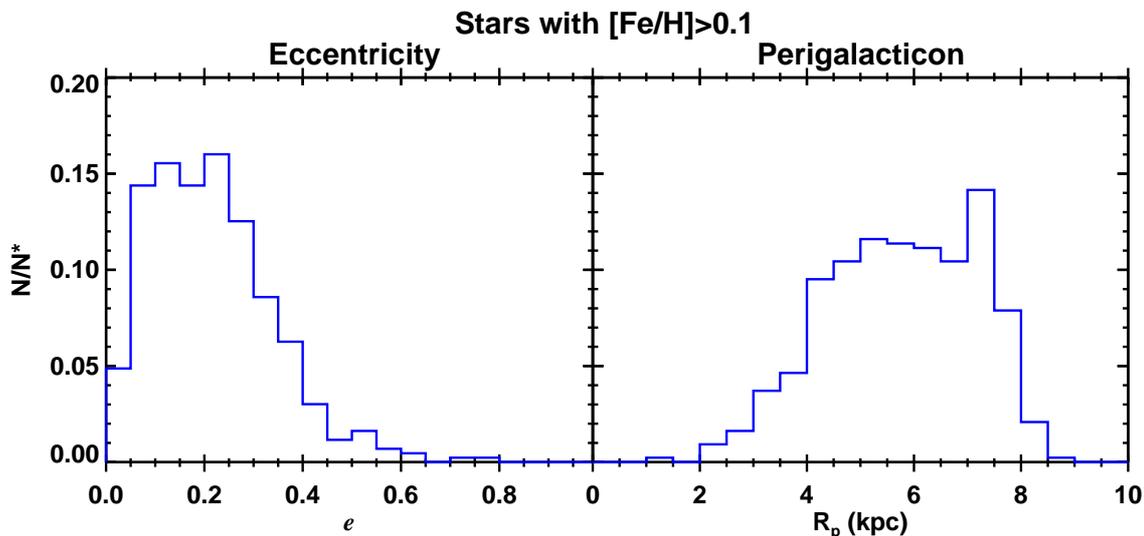}
\caption{\textbf{Left:} Distributions of orbital eccentricities for stars with [Fe/H]$>0.1$. \textbf{Right:} The distribution of perigalacticons for stars with [Fe/H]$>0.1$.}
\label{orbits}
\end{figure*}

 We used stars with high metallicities ([Fe/H]>0.1) to get a lower limit on the fraction of stars that have undergone radial migration. These stars have metallicities higher than the local ISM ($\sim$solar), which have likely undergone some form of radial mixing to be observed in the extended solar neighborhood. We find that half of the metal-rich stars have eccentricities $e<0.2$ or are on roughly circular orbits, as shown in the left panel of Fig. \ref{orbits}. Furthermore, one-quarter of the metal-rich stars have perigalacticons $R_{p}>7$ kpc (right panel, Fig. \ref{orbits}). For these stars, migration is likely an important process, as their orbits do not reach the radii at which they were likely to be born based on their high metallicities. For example, a star with [Fe/H]$\sim0.2$ likely formed inside $\sim5-6$ kpc. The fraction of stars with $R_p>7$ kpc is relatively consistent for stars with significantly super-solar metallicities ([Fe/H]$>0.25$), as noted in Table 2.1. This implies that this result is not due to the effects of the radial gradient; i.e., we are not simply observing the most metal-rich stars forming in situ at $R\sim7$ kpc (ISM metallicity $\sim$0.1), as the the relative fraction of metal-rich stars with $R_p>7$ kpc is roughly constant even with extremely strict definitions of metal-rich populations.

\section{Discussion}

Our results for the distribution of eccentricities of stars that are likely to be members of the thick disk stellar populations (red stars in Fig. \ref{survey}) in iDR4 are similar to those of past GES data releases \citep{Recio-Blanco2014}, but with a large increase in sample size. We find that the orbital eccentricity of the thick disk decreases as a function of metallicity, as observed previously by \citet{Recio-Blanco2014}. The median eccentricity is very large for the most metal-poor thick disk populations, $\sim0.35$ and gradually decreases to eccentricities that approach the thin disk distribution with median eccentricities of $\sim0.2$ for the more metal-rich ([Fe/H] larger than solar) thick disk stars. This highlights the potential danger of using kinematics to attempt to separate the thin and thick disk because the kinematics for the thick disk is a strong function of metallicity and appears similar to thin disk populations for the more metal-rich portion of the thick disk. 

The RAVE observations show a clear decrease in velocity dispersion for the highest [Mg/Fe] metal-poor populations \citep{Minchev2014} in all velocity components. This decrease is also seen in simulations from \citet{Minchev2013} and is caused by a merger event early in the simulation ($z\sim2-3$). In the simulation, the merger drives enhanced radial migration, which preferentially affects the kinematically coolest populations (see, e.g., \citealt{Solway2012,VeraCiro2014}). This causes large numbers of kinematically cooler stars from the inner disk to migrate to larger radii, which is then observed as a decrease in velocity dispersion for the older, higher [Mg/Fe] stellar populations. This signature is observed both in the simulation and solar neighborhood with RAVE. Further, this enhanced migration of kinematically cooler populations causes an increase in the observed rotational velocity of the highest [Mg/Fe] stars, which also translates to stars having slightly less eccentric orbits and larger perigalacticons. 

Observations from the GES survey for this result are less obvious, as the dip in velocity dispersions is much less significant in the iDR4 observations compared to RAVE. The velocity dispersion decreases by $\sim10$ km s$^{-1}$ for the most metal-poor, alpha-rich populations in the vertical and radial components, but not the rotational component. This behavior was found previously in GES by \citet{Guiglion2015} and the increased sample size of iDR4 only slightly alleviates this issue. It is possible this is due to sampling issues; GES is focused on stars away from the midplane, where the vertical velocities are close to their minimum values. This is also observed in the SEGUE results presented by \citet{Minchev2014}, which is also focused above the midplane, where the most metal-poor populations show a much smaller decrease in vertical velocity dispersion compared to the RAVE results and also show no reversal in the rotational velocity dispersion, similar to GES observations. 

The only major difference compared to the iDR2 results from \citet{Guiglion2015} is a slightly higher radial velocity dispersion for low-[Mg/Fe] populations, which is caused by the addition of giants and other more distant stars with larger systematic velocity uncertainties in the iDR2 results compared to the iDR4 results presented here. Qualitatively, the results between DR2 and iDR4 agree very well.

The largest potential source of tension for enhanced migration driven by a merger is the orbital properties for the metal-poor populations. The [Mg/Fe]-poor populations make up the thin disk and have fairly circular orbits with eccentricities of $e\sim0.15$. This is in contrast to the [Mg/Fe]-enhanced populations, which show larger eccentricities on average that are more similar to those of the thick disk stars in Fig. \ref{thickecc}. However, migration can occur even for kinematically hot populations because it is simply more efficient for those with cool kinematics (e.g., \citealt{VeraCiro2014}). Furthermore, if the bulk of the disk at early times was kinematically hot in general, even those populations that migrate do not have dramatically cooler kinematics compared to thin disk stars today. The key point is that if the higher \af populations migrated, their kinematics are cooler than those lower \af stars that arrive in the solar neighborhood via blurring only. Our observations from iDR4 prove to be somewhat inconclusive for this result: we find small decreases in velocity dispersion, eccentricity observed for many high-[Mg/Fe] populations relative to the lower [Mg/Fe] stars of similar metallicity, but the observed trends are not as clear as the RAVE velocity dispersions. 

Stars with [Fe/H]$>0.1$ are significantly metal-enhanced compared to the local ISM, which is $\sim$solar metallicity. It is therefore likely that these stars did not form in the solar neighborhood, but are observed here due to secular processes such as blurring or migration. The relative importance of secular processes is currently a matter of ongoing debate. We used the metal-rich populations observed in GES to attempt to constrain the fraction of stars that have likely undergone significant radial migration. The orbital eccentricity is a useful quantity to estimate the relative importance of these processes on this metal-rich population and potentially the Milky Way as a whole. We find that at least one-fifth of these stars are likely to have undergone radial migration. This is a lower limit; it is possible that the fraction of migrated stars is (much) larger. However, we can only say for certain that slightly more than 20\% of the metal-rich population do not reach the orbital radii at which they were likely born. 

We have no direct measure of the distance these stars have migrated. The radial gradient for the bulk stellar populations in the inner Galaxy appears to be fairly flat \citep{Hayden2014}. Moreover, the MDF is strongly peaked at high metallicities both in the very inner disk ($3<R<5$ kpc) and also at larger radii ($5<R<7$ kpc, \citealt{Hayden2015}). It is therefore not required that many of the metal-rich stars are likely to have migrated to travel extremely large distances; it is possible that these stars migrated only 1 or 2 kpc to their present locations. In this sense, our results are not necessarily in conflict with those of \citet{Haywood2016,Dimatteo2016}, who argue that no large-scale migration of stellar populations of many kiloparsecs has occurred in the Milky Way.

\section{Conclusions}

We analyzed the chemical and kinematic properties of stars in the iDR4 release of the GES to look for evidence for past mergers and the relative fraction of stars that might have undergone radial migration. We find that the eccentricities of thick disk stars decrease as metallicities increase and that the most metal-rich stars have eccentricities similar to those of the thin disk. We find that observations in iDR4 are not inconsistent with previous results from RAVE showing a decrease in velocity dispersion for the highest [Mg/Fe] (oldest) stellar populations, potentially caused by a merger that drives enhanced radial migration. However, these results are less obvious in the iDR4 observations. We also find that 20\% of very metal-rich ([Fe/H]$>0.1$) stars have perigalacticons larger than 7 kpc, meaning their orbits never reach the inner Galaxy where they had to have formed. These stars are likely to have migrated from their original birth radii to their present location in the extended solar neighborhood. 

There has been much debate about the relative impact of radial migration and blurring on the structure and evolution of the Milky Way. Our observations from the GES survey have shown that a large numbers of stars are likely to have undergone migration of some kind. We note that our results cannot constrain the distance migrated, and merely that these stars never reach radii at which they likely formed based on their metallicities. We do not constrain how far the stars have migrated; it is possible that many of these stars only have to migrate 1 or 2 kpc to reach their present orbits. We merely find that migration is needed to explain the presence of very metal-rich stars in the solar neighborhood that have circular orbits. 

Our results show the power that large-scale surveys and precision kinematic measurements have to unravel the past history of the Galaxy. With the upcoming data releases from Gaia \citep{Brown2016}, proper motions and parallax measurements will be available for millions of stars throughout the Milky Way. The sample size and accuracy of the measurements used in this paper can be improved by an order of magnitude with these observations and further aid our ability to discover the past history of the Milky Way. In addition, improvements in models and simulations will provide additional points of comparison with observations. Our observations provide only a lower limit on the fraction of stars that undergo migration and cannot constrain the radial extent that stars have migrated. Models and simulations are key to providing additional insight into the evolution of the Galaxy, such as those developed by \citet{Minchev2013}.

\begin{acknowledgements}
The authors thank Ivan Minchev for a useful discussion and the anonymous referee for useful comments. MRH and ARB received financial support from ANR, reference 14-CE33-0014-01. SF and TB acknowledge support from the project grant “The New Milky Way” from the Knut and Alice Wallenberg Foundation. Based on data products from observations made with ESO Telescopes at the La Silla Paranal Observatory under program ID 188.B-3002. These data products have been processed by the Cambridge Astronomy Survey Unit (CASU) at the Institute of Astronomy, University of Cambridge, and by the FLAMES/UVES reduction team at INAF/Osservatorio Astrofisico di Arcetri. These data have been obtained from the Gaia-ESO Survey Data Archive, prepared and hosted by the Wide Field Astronomy Unit, Institute for Astronomy, University of Edinburgh, which is funded by the UK Science and Technology Facilities Council. This work was partly supported by the European Union FP7 programme through ERC grant number 320360 and by the Leverhulme Trust through grant RPG-2012-541. We acknowledge the support from INAF and Ministero dell’ Istruzione, dell’ Università’ e della Ricerca (MIUR) in the form of the grant "Premiale VLT 2012". The results presented here benefit from discussions held during the Gaia-ESO workshops and conferences supported by the ESF (European Science Foundation) through the GREAT Research Network Programme. This research has made use of the SIMBAD database, operated at CDS, Strasbourg, France and NASA’s Astrophysics Data System.

\end{acknowledgements}

\bibliographystyle{aa}
\bibliography{ref}

\begin{thebibliography}{47}
\expandafter\ifx\csname natexlab\endcsname\relax\def\natexlab#1{#1}\fi

\bibitem[{Adibekyan {et~al.}(2012)Adibekyan, Sousa, Santos, {Delgado Mena},
  {Gonz{\'{a}}lez Hern{\'{a}}ndez}, Israelian, Mayor, \&
  Khachatryan}]{Adibekyan2012}
Adibekyan, V.~Z., Sousa, S.~G., Santos, N.~C., {et~al.} 2012, A{\&}A, 545, A32

\bibitem[{{Allende Prieto} {et~al.}(2006){Allende Prieto}, Beers, Wilhelm,
  Newberg, Rockosi, Yanny, \& Lee}]{AllendePrieto2005}
{Allende Prieto}, C., Beers, T.~C., Wilhelm, R., {et~al.} 2006, ApJ, 636, 804

\bibitem[{Bensby {et~al.}(2003)Bensby, Feltzing, \&
  Lundstr{\"{o}}m}]{Bensby2003}
Bensby, T., Feltzing, S., \& Lundstr{\"{o}}m, I. 2003, A{\&}A, 410, 527

\bibitem[{Bensby {et~al.}(2014)Bensby, Feltzing, \& Oey}]{Bensby2014}
Bensby, T., Feltzing, S., \& Oey, M.~S. 2014, A{\&}A, 562, A71

\bibitem[{Bensby {et~al.}(2007)Bensby, Zenn, Oey, \& Feltzing}]{Bensby2007}
Bensby, T., Zenn, A.~R., Oey, M.~S., \& Feltzing, S. 2007, ApJ, 663, L13

\bibitem[{Bird {et~al.}(2012)Bird, Kazantzidis, \& Weinberg}]{Bird2012}
Bird, J.~C., Kazantzidis, S., \& Weinberg, D.~H. 2012, MNRAS, 420, 913

\bibitem[{Bovy(2015)}]{Bovy2015a}
Bovy, J. 2015, ApJS, 216, 29

\bibitem[{Bressan {et~al.}(2012)Bressan, Marigo, Girardi, Salasnich, {Dal
  Cero}, Rubele, \& Nanni}]{Bressan2012}
Bressan, A., Marigo, P., Girardi, L., {et~al.} 2012, MNRAS, 427, 127

\bibitem[{Brown {et~al.}(2016)Brown, Vallenari, Prusti, de~Bruijne, Mignard,
  Drimmel, Co-authors, Bailer-Jones, Bastian, Biermann, Evans, Eyer, Jansen,
  Jordi, Katz, Klioner, Lammers, Lindegren, Luri, O'Mullane, Panem, Pourbaix,
  Randich, Sartoretti, Siddiqui, Soubiran, Valette, van Leeuwen, Walton, Aerts,
  Arenou, Cropper, H{\o}g, Lattanzi, Grebel, Holland, Huc, Passot, Perryman,
  Bramante, Cacciari, Casta{\~{n}}eda, Chaoul, Cheek, {De Angeli}, Fabricius,
  Guerra, Hern{\'{a}}ndez, Jean-Antoine-Piccolo, Masana, Messineo, Mowlavi,
  Nienartowicz, Ord{\'{o}}{\~{n}}ez-Blanco, Panuzzo, Portell, Richards, Riello,
  Seabroke, Tanga, Th{\'{e}}venin, Torra, Els, Gracia-Abril, Comoretto,
  Garcia-Reinaldos, Lock, Mercier, Altmann, Andrae, Astraatmadja,
  Bellas-Velidis, Benson, Berthier, Blomme, Busso, Carry, Cellino, Clementini,
  Cowell, Creevey, Cuypers, Davidson, {De Ridder}, de~Torres, Delchambre,
  Dell'Oro, Ducourant, Fr{\'{e}}mat, Garc{\'{i}}a-Torres, Gosset, Halbwachs,
  Hambly, Harrison, Hauser, Hestroffer, Hodgkin, Huckle, Hutton, Jasniewicz,
  Jordan, Kontizas, Korn, Lanzafame, Manteiga, Moitinho, Muinonen, Osinde,
  Pancino, Pauwels, Petit, Recio-Blanco, Robin, Sarro, Siopis, Smith, Smith,
  Sozzetti, Thuillot, van Reeven, Viala, Abbas, {Abreu Aramburu}, Accart,
  Aguado, Allan, Allasia, Altavilla, {\'{A}}lvarez, Alves, Anderson, Andrei,
  {Anglada Varela}, Antiche, Antoja, Ant{\'{o}}n, Arcay, Bach, Baker,
  Balaguer-N{\'{u}}{\~{n}}ez, Barache, Barata, Barbier, Barblan, {Barrado y
  Navascu{\'{e}}s}, Barros, Barstow, Becciani, Bellazzini, {Bello
  Garc{\'{i}}a}, Belokurov, Bendjoya, Berihuete, Bianchi, Bienaym{\'{e}},
  Billebaud, Blagorodnova, Blanco-Cuaresma, Boch, Bombrun, Borrachero,
  Bouquillon, Bourda, Bouy, Bragaglia, Breddels, Brouillet, Br{\"{u}}semeister,
  Bucciarelli, Burgess, Burgon, Burlacu, Busonero, Buzzi, Caffau, Cambras,
  Campbell, Cancelliere, Cantat-Gaudin, Carlucci, Carrasco, Castellani,
  Charlot, Charnas, Chiavassa, Clotet, Cocozza, Collins, Costigan, Crifo,
  Cross, Crosta, Crowley, Dafonte, Damerdji, Dapergolas, David, David, {De
  Cat}, de~Felice, de~Laverny, {De Luise}, {De March}, de~Martino, de~Souza,
  Debosscher, del Pozo, Delbo, Delgado, Delgado, {Di Matteo}, Diakite,
  Distefano, Dolding, {Dos Anjos}, Drazinos, Duran, Dzigan, Edvardsson, Enke,
  Evans, {Eynard Bontemps}, Fabre, Fabrizio, Faigler, Falc{\~{a}}o,
  {Farr{\`{a}}s Casas}, Federici, Fedorets, Fern{\'{a}}ndez-Hern{\'{a}}ndez,
  Fernique, Fienga, Figueras, Filippi, Findeisen, Fonti, Fouesneau, Fraile,
  Fraser, Fuchs, Gai, Galleti, Galluccio, Garabato, Garc{\'{i}}a-Sedano,
  Garofalo, Garralda, Gavras, Gerssen, Geyer, Gilmore, Girona, Giuffrida,
  Gomes, Gonz{\'{a}}lez-Marcos, Gonz{\'{a}}lez-N{\'{u}}{\~{n}}ez,
  Gonz{\'{a}}lez-Vidal, Granvik, Guerrier, Guillout, Guiraud, G{\'{u}}rpide,
  Guti{\'{e}}rrez-S{\'{a}}nchez, Guy, Haigron, Hatzidimitriou, Haywood, Heiter,
  Helmi, Hobbs, Hofmann, Holl, Holland, Hunt, Hypki, Icardi, Irwin, {Jevardat
  de Fombelle}, Jofr{\'{e}}, Jonker, Jorissen, Julbe, Karampelas, Kochoska,
  Kohley, Kolenberg, Kontizas, Koposov, Kordopatis, Koubsky, Krone-Martins,
  Kudryashova, Kull, Bachchan, Lacoste-Seris, Lanza, Lavigne, {Le
  Poncin-Lafitte}, Lebreton, Lebzelter, Leccia, Leclerc, Lecoeur-Taibi,
  Lemaitre, Lenhardt, Leroux, Liao, Licata, Lindstr{\o}m, Lister, Livanou,
  Lobel, L{\"{o}}ffler, L{\'{o}}pez, Lorenz, MacDonald, {Magalh{\~{a}}es
  Fernandes}, Managau, Mann, Mantelet, Marchal, Marchant, Marconi, Marinoni,
  Marrese, Marschalk{\'{o}}, Marshall, Mart{\'{i}}n-Fleitas, Martino, Mary,
  Matijevi{\v{c}}, Mazeh, McMillan, Messina, Michalik, Millar, Miranda, Molina,
  Molinaro, Molinaro, Moln{\'{a}}r, Moniez, Montegriffo, Mor, Mora, Morbidelli,
  Morel, Morgenthaler, Morris, Mulone, Muraveva, Musella, Narbonne, Nelemans,
  Nicastro, Noval, Ord{\'{e}}novic, Ordieres-Mer{\'{e}}, Osborne, Pagani,
  Pagano, Pailler, Palacin, Palaversa, Parsons, Pecoraro, Pedrosa,
  Pentik{\"{a}}inen, Pichon, Piersimoni, Pineau, Plachy, Plum, Poujoulet,
  Pr{\v{s}}a, Pulone, Ragaini, Rago, Rambaux, Ramos-Lerate, Ranalli, Rauw,
  Read, Regibo, Reyl{\'{e}}, Ribeiro, Rimoldini, Ripepi, Riva, Rixon, Roelens,
  Romero-G{\'{o}}mez, Rowell, Royer, Ruiz-Dern, Sadowski, {Sagrist{\`{a}}
  Sell{\'{e}}s}, Sahlmann, Salgado, Salguero, Sarasso, Savietto, Schultheis,
  Sciacca, Segol, Segovia, Segransan, Shih, Smareglia, Smart, Solano, Solitro,
  Sordo, {Soria Nieto}, Souchay, Spagna, Spoto, Stampa, Steele,
  Steidelm{\"{u}}ller, Stephenson, Stoev, Suess, S{\"{u}}veges, Surdej,
  Szabados, Szegedi-Elek, Tapiador, Taris, Tauran, Taylor, Teixeira, Terrett,
  Tingley, Trager, Turon, Ulla, Utrilla, Valentini, van Elteren, {Van
  Hemelryck}, van Leeuwen, Varadi, Vecchiato, Veljanoski, Via, Vicente, Vogt,
  Voss, Votruba, Voutsinas, Walmsley, Weiler, Weingrill, Wevers, Wyrzykowski,
  Yoldas, {\v{Z}}erjal, Zucker, Zurbach, Zwitter, Alecu, Allen, {Allende
  Prieto}, Amorim, Anglada-Escud{\'{e}}, Arsenijevic, Azaz, Balm, Beck,
  Bernstein, Bigot, Bijaoui, Blasco, Bonfigli, Bono, Boudreault, Bressan,
  Brown, Brunet, Bunclark, Buonanno, Butkevich, Carret, Carrion, Chemin,
  Ch{\'{e}}reau, Corcione, Darmigny, de~Boer, de~Teodoro, de~Zeeuw, {Delle
  Luche}, Domingues, Dubath, Fodor, Fr{\'{e}}zouls, Fries, Fustes, Fyfe,
  Gallardo, Gallegos, Gardiol, Gebran, Gomboc, G{\'{o}}mez, Grux, Gueguen,
  Heyrovsky, Hoar, Iannicola, {Isasi Parache}, Janotto, Joliet, Jonckheere,
  Keil, Kim, Klagyivik, Klar, Knude, Kochukhov, Kolka, Kos, Kutka, Lainey,
  LeBouquin, Liu, Loreggia, Makarov, Marseille, Martayan, Martinez-Rubi,
  Massart, Meynadier, Mignot, Munari, Nguyen, Nordlander, Ocvirk, O'Flaherty,
  {Olias Sanz}, Ortiz, Osorio, Oszkiewicz, Ouzounis, Palmer, Park, Pasquato,
  Peltzer, Peralta, P{\'{e}}turaud, Pieniluoma, Pigozzi, Poels, Prat,
  Prod'homme, Raison, Rebordao, Risquez, Rocca-Volmerange, Rosen, Ruiz-Fuertes,
  Russo, Sembay, {Serraller Vizcaino}, Short, Siebert, Silva, Sinachopoulos,
  Slezak, Soffel, Sosnowska, Strai{\v{z}}ys, ter Linden, Terrell, Theil, Tiede,
  Troisi, Tsalmantza, Tur, Vaccari, Vachier, Valles, {Van Hamme}, Veltz,
  Virtanen, Wallut, Wichmann, Wilkinson, Ziaeepour, \& Zschocke}]{Brown2016}
Brown, A. G.~A., Vallenari, A., Prusti, T., {et~al.} 2016, A{\&}A, 595, A2

\bibitem[{Chabrier(2003)}]{Chabrier2003}
Chabrier, G. 2003, PASP, 115, 763

\bibitem[{Coşkunoğlu {et~al.}(2011)Coşkunoğlu, Ak, Bilir, Karaali, Yaz,
  Gilmore, Seabroke, Bienaym{\'{e}}, Bland-Hawthorn, Campbell, Freeman, Gibson,
  Grebel, Munari, Navarro, Parker, Siebert, Siviero, Steinmetz, Watson, Wyse,
  \& Zwitter}]{Coskunoglu2011}
Coşkunoğlu, B., Ak, S., Bilir, S., {et~al.} 2011, MNRAS, 412, 1237

\bibitem[{{Di Matteo}(2016)}]{Dimatteo2016}
{Di Matteo}, P. 2016, PASA, 33, e027

\bibitem[{D'Onghia {et~al.}(2015)D'Onghia, Madau, Vera-Ciro, Quillen, \&
  Hernquist}]{Donghia2016}
D'Onghia, E., Madau, P., Vera-Ciro, C., Quillen, A., \& Hernquist, L. 2015,
  ApJ, 823

\bibitem[{Freeman(2010)}]{Freeman2010}
Freeman, K.~C. 2010, {Galaxies and their Masks}, ed. D.~L. Block, K.~C.
  Freeman, \& I.~Puerari (New York, NY: Springer New York)

\bibitem[{Fuhrmann(1998)}]{Fuhrmann1998}
Fuhrmann, K. 1998, A{\&}A, 338, 161

\bibitem[{Gilmore {et~al.}(2012)Gilmore, Randich, Asplund, Binney, Bonifacio,
  Drew, Feltzing, Ferguson, Jeffries, Micela, Negueruela, Prusti, Rix,
  Vallenari, Alfaro, Allende-Prieto, Babusiaux, Bensby, Blomme, Bragaglia,
  Flaccomio, Fran{\c{c}}ois, Irwin, Koposov, Korn, Lanzafame, Pancino, Paunzen,
  Recio-Blanco, Sacco, Smiljanic, {Van Eck}, Walton, Aden, Aerts, Affer,
  Alcala, Altavilla, Alves, Antoja, Arenou, Argiroffi, {Asensio Ramos},
  Bailer-Jones, Balaguer-Nunez, Bayo, Barbuy, Barisevicius, {Barrado y
  Navascues}, Battistini, {Bellas Velidis}, Bellazzini, Belokurov, Bergemann,
  Bertelli, Biazzo, Bienayme, Bland-Hawthorn, Boeche, Bonito, Boudreault,
  Bouvier, Brandao, Brown, de~Bruijne, Burleigh, Caballero, Caffau, Calura,
  Capuzzo-Dolcetta, Caramazza, Carraro, Casagrande, Casewell, Chapman,
  Chiappini, Chorniy, Christlieb, Cignoni, Cocozza, Colless, Collet, Collins,
  Correnti, Covino, Crnojevic, Cropper, Cunha, Damiani, David, Delgado, Duffau,
  Edvardsson, Eldridge, Enke, Eriksson, Evans, Eyer, Famaey, Fellhauer,
  Ferreras, Figueras, Fiorentino, Flynn, Folha, Franciosini, Frasca, Freeman,
  Fremat, Friel, Gaensicke, Gameiro, Garzon, Geier, Geisler, Gerhard, Gibson,
  Gomboc, Gomez, Gonzalez-Fernandez, {Gonzalez Hernandez}, Gosset, Grebel,
  Greimel, Groenewegen, Grundahl, Guarcello, Gustafsson, Hadrava,
  Hatzidimitriou, Hambly, Hammersley, Hansen, Haywood, Heber, Heiter, Held,
  Helmi, Hensler, Herrero, Hill, Hodgkin, Huelamo, Huxor, Ibata, Jackson,
  de~Jong, Jonker, Jordan, Jordi, Jorissen, Katz, Kawata, Keller, Kharchenko,
  Klement, Klutsch, Knude, Koch, Kochukhov, Kontizas, Koubsky, Lallement,
  de~Laverny, van Leeuwen, Lemasle, Lewis, Lind, Lindstrom, Lobel, {Lopez
  Santiago}, Lucas, Ludwig, Lueftinger, Magrini, {Maiz Apellaniz}, Maldonado,
  Marconi, Marino, Martayan, Martinez-Valpuesta, Matijevic, McMahon, Messina,
  Meyer, Miglio, Mikolaitis, Minchev, Minniti, Moitinho, Momany, Monaco,
  Montalto, Monteiro, Monier, Montes, Mora, Moraux, Morel, Mowlavi,
  Mucciarelli, Munari, Napiwotzki, Nardetto, Naylor, Naze, Nelemans, Okamoto,
  Ortolani, Pace, Palla, Palous, Parker, Penarrubia, Pillitteri, Piotto,
  Posbic, Prisinzano, Puzeras, Quirrenbach, Ragaini, Read, Read, Reyle, {De
  Ridder}, Robichon, Robin, Roeser, Romano, Royer, Ruchti, Ruzicka, Ryan, Ryde,
  Santos, {Sanz Forcada}, {Sarro Baro}, Sbordone, Schilbach, Schmeja, Schnurr,
  Schoenrich, Scholz, Seabroke, Sharma, {De Silva}, Smith, Solano, Sordo,
  Soubiran, Sousa, Spagna, Steffen, Steinmetz, Stelzer, Stempels, Tabernero,
  Tautvaisiene, Thevenin, Torra, Tosi, Tolstoy, Turon, Walker, Wambsganss,
  Worley, Venn, Vink, Wyse, Zaggia, Zeilinger, Zoccali, Zorec, Zucker, Zwitter,
  \& Team}]{Gilmore2012}
Gilmore, G., Randich, S., Asplund, M., {et~al.} 2012, The Messenger, 147, 25

\bibitem[{Gilmore \& Reid(1983)}]{Gilmore1983}
Gilmore, G. \& Reid, N. 1983, MNRAS, 202, 1025

\bibitem[{Godwin \& Lynden-Bell(1987)}]{Godwin1987}
Godwin, P.~J. \& Lynden-Bell, D. 1987, MNRAS, 229, 7P

\bibitem[{Guiglion {et~al.}(2015)Guiglion, Recio-Blanco, de~Laverny,
  Kordopatis, Hill, Mikolaitis, Minchev, Chiappini, Wyse, Gilmore, Randich,
  Feltzing, Bensby, Flaccomio, Koposov, Pancino, Bayo, Costado, Franciosini,
  Hourihane, Jofr{\'{e}}, Lardo, Lewis, Lind, Magrini, Morbidelli, Sacco,
  Ruchti, Worley, \& Zaggia}]{Guiglion2015}
Guiglion, G., Recio-Blanco, A., de~Laverny, P., {et~al.} 2015, A{\&}A, 583, A91

\bibitem[{Hayden {et~al.}(2015)Hayden, Bovy, Holtzman, Nidever, Bird, Weinberg,
  Andrews, Majewski, Prieto, Anders, Beers, Bizyaev, Chiappini, Cunha,
  Frinchaboy, Garc{\'{i}}a-Her{\'{n}}andez, {Garc{\'{i}}a P{\'{e}}rez},
  Girardi, Harding, Hearty, Johnson, M{\'{e}}sz{\'{a}}ros, Minchev, O'Connell,
  Pan, Robin, Schiavon, Schneider, Schultheis, Shetrone, Skrutskie, Steinmetz,
  Smith, Wilson, Zamora, \& Zasowski}]{Hayden2015}
Hayden, M.~R., Bovy, J., Holtzman, J.~A., {et~al.} 2015, ApJ, 808, 132

\bibitem[{Hayden {et~al.}(2014)Hayden, Holtzman, Bovy, Majewski, Johnson,
  {Allende Prieto}, Beers, Cunha, Frinchaboy, {Garc{\'{i}}a P{\'{e}}rez},
  Girardi, Hearty, Lee, Nidever, Schiavon, Schlesinger, Schneider, Schultheis,
  Shetrone, Smith, Zasowski, Bizyaev, Feuillet, Hasselquist, Kinemuchi,
  Malanushenko, Malanushenko, O'Connell, Pan, \& Stassun}]{Hayden2014}
Hayden, M.~R., Holtzman, J.~A., Bovy, J., {et~al.} 2014, AJ, 147, 116

\bibitem[{Haywood {et~al.}(2013)Haywood, {Di Matteo}, Lehnert, Katz, \&
  G{\'{o}}mez}]{Haywood2013}
Haywood, M., {Di Matteo}, P., Lehnert, M.~D., Katz, D., \& G{\'{o}}mez, A.
  2013, A{\&}A, 560, A109

\bibitem[{Haywood {et~al.}(2016)Haywood, Lehnert, {Di Matteo}, Snaith,
  Schultheis, Katz, \& Gomez}]{Haywood2016}
Haywood, M., Lehnert, M.~D., {Di Matteo}, P., {et~al.} 2016, A{\&}A, 589, A66

\bibitem[{Hopkins {et~al.}(2008)Hopkins, Hernquist, Cox, Younger, \&
  Besla}]{Hopkins2008}
Hopkins, P.~F., Hernquist, L., Cox, T.~J., Younger, J.~D., \& Besla, G. 2008,
  ApJ, 688

\bibitem[{Kordopatis {et~al.}(2011)Kordopatis, Recio-Blanco, {De Laverny},
  Gilmore, Hill, Wyse, Helmi, Bijaoui, Zoccali, \&
  Bienaym{\'{e}}}]{Kordopatis2011}
Kordopatis, G., Recio-Blanco, A., {De Laverny}, P., {et~al.} 2011, A{\&}A, 535,
  A107

\bibitem[{Kordopatis {et~al.}(2015)Kordopatis, Wyse, Gilmore, Recio-Blanco,
  de~Laverny, Hill, Adibekyan, Heiter, Minchev, Famaey, Bensby, Feltzing,
  Guiglion, Korn, Mikolaitis, Schultheis, Vallenari, Bayo, Carraro, Flaccomio,
  Franciosini, Hourihane, Jofr{\'{e}}, Koposov, Lardo, Lewis, Lind, Magrini,
  Morbidelli, Pancino, Randich, Sacco, Worley, \& Zaggia}]{Kordopatis2015}
Kordopatis, G., Wyse, R. F.~G., Gilmore, G., {et~al.} 2015, A{\&}A, 582, A122

\bibitem[{Lee {et~al.}(2011)Lee, Beers, An, Ivezi{\'{c}}, Just, Rockosi,
  Morrison, Johnson, Sch{\"{o}}nrich, Bird, Yanny, Harding, \&
  Rocha-Pinto}]{Lee2011}
Lee, Y.~S., Beers, T.~C., An, D., {et~al.} 2011, ApJ, 738, 187

\bibitem[{Loebman {et~al.}(2016)Loebman, Debattista, Nidever, Hayden, Holtzman,
  Clarke, Ro{\v{s}}kar, \& Valluri}]{Loebman2016}
Loebman, S.~R., Debattista, V.~P., Nidever, D.~L., {et~al.} 2016, ApJ, 818, L6

\bibitem[{Loebman {et~al.}(2011)Loebman, Ro{\v{s}}kar, Debattista,
  Ivezi{\'{c}}, Quinn, \& Wadsley}]{Loebman2011}
Loebman, S.~R., Ro{\v{s}}kar, R., Debattista, V.~P., {et~al.} 2011, ApJ, 737, 8

\bibitem[{Majewski {et~al.}(2015)Majewski, Schiavon, Frinchaboy, Prieto,
  Barkhouser, Bizyaev, Blank, Brunner, Burton, Carrera, Chojnowski, Cunha,
  Epstein, Fitzgerald, Perez, Hearty, Henderson, Holtzman, Johnson, Lam,
  Lawler, Maseman, Meszaros, Nelson, Nguyen, Nidever, Pinsonneault, Shetrone,
  Smee, Smith, Stolberg, Skrutskie, Walker, Wilson, Zasowski, Anders, Basu,
  Beland, Blanton, Bovy, Brownstein, Carlberg, Chaplin, Chiappini, Eisenstein,
  Elsworth, Feuillet, Fleming, Galbraith-Frew, Garcia, Garcia-Hernandez,
  Gillespie, Girardi, Gunn, Hasselquist, Hayden, Hekker, Ivans, Kinemuchi,
  Klaene, Mahadevan, Mathur, Mosser, Muna, Munn, Nichol, O'Connell, Robin,
  Rocha-Pinto, Schultheis, Serenelli, Shane, Aguirre, Sobeck, Thompson, Troup,
  Weinberg, \& Zamora}]{Majewski2015}
Majewski, S.~R., Schiavon, R.~P., Frinchaboy, P.~M., {et~al.} 2015, eprint
  arXiv:1509.05420

\bibitem[{Minchev {et~al.}(2013)Minchev, Chiappini, \& Martig}]{Minchev2013}
Minchev, I., Chiappini, C., \& Martig, M. 2013, A{\&}A, 558, A9

\bibitem[{Minchev {et~al.}(2014)Minchev, Chiappini, Martig, Steinmetz, de~Jong,
  Boeche, Scannapieco, Zwitter, Wyse, Binney, Bland-Hawthorn, Bienaym{\'{e}},
  Famaey, Freeman, Gibson, Grebel, Gilmore, Helmi, Kordopatis, Lee, Munari,
  Navarro, Parker, Quillen, Reid, Siebert, Siviero, Seabroke, Watson, \&
  Williams}]{Minchev2014}
Minchev, I., Chiappini, C., Martig, M., {et~al.} 2014, ApJ, 781, L20

\bibitem[{Navarro {et~al.}(1997)Navarro, Frenk, \& White}]{Navarro1997}
Navarro, J.~F., Frenk, C.~S., \& White, S. D.~M. 1997, ApJ, 490, 493

\bibitem[{Piskunov \& Valenti(2016)}]{Piskunov2016}
Piskunov, N. \& Valenti, J.~A. 2016, eprint arXiv:1606.06073

\bibitem[{Pryor \& Meylan(1993)}]{Pryor1993}
Pryor, C. \& Meylan, G. 1993, Structure and Dynamics of Globular Clusters.
  Proceedings of a Workshop held in Berkeley, 50

\bibitem[{Recio-Blanco {et~al.}(2006)Recio-Blanco, Bijaoui, \& {De
  Laverny}}]{Recio-Blanco2006}
Recio-Blanco, A., Bijaoui, A., \& {De Laverny}, P. 2006, MNRAS, 370, 141

\bibitem[{Recio-Blanco {et~al.}(2014)Recio-Blanco, de~Laverny, Kordopatis,
  Helmi, Hill, Gilmore, Wyse, Adibekyan, Randich, Asplund, Feltzing, Jeffries,
  Micela, Vallenari, Alfaro, Prieto, Bensby, Bragaglia, Flaccomio, Koposov,
  Korn, Lanzafame, Pancino, Smiljanic, Jackson, Lewis, Magrini, Morbidelli,
  Prinsinzano, Sacco, Worley, Hourihane, Bergemann, Costado, Heiter, Joffre,
  Lardo, Lind, \& Maiorca}]{Recio-Blanco2014}
Recio-Blanco, A., de~Laverny, P., Kordopatis, G., {et~al.} 2014, A{\&}A, 567

\bibitem[{Reddy {et~al.}(2006)Reddy, Lambert, \& Prieto}]{Reddy2006}
Reddy, B.~E., Lambert, D.~L., \& Prieto, C.~A. 2006, MNRAS, 367, 1329

\bibitem[{Roeser {et~al.}(2010)Roeser, Demleitner, \& Schilbach}]{Roeser2010}
Roeser, S., Demleitner, M., \& Schilbach, E. 2010, AJ, 139, 2440

\bibitem[{Sch{\"{o}}nrich \& Binney(2009)}]{Schonrich2009}
Sch{\"{o}}nrich, R. \& Binney, J. 2009, MNRAS, 399, 1145

\bibitem[{Sellwood \& Binney(2002)}]{Sellwood2002}
Sellwood, J.~A. \& Binney, J.~J. 2002, MNRAS, 336, 785

\bibitem[{Solway {et~al.}(2012)Solway, Sellwood, \&
  Sch{\"{o}}nrich}]{Solway2012}
Solway, M., Sellwood, J.~A., \& Sch{\"{o}}nrich, R. 2012, MNRAS, 422, 1363

\bibitem[{Steinmetz {et~al.}(2006)Steinmetz, Zwitter, Siebert, Watson, Freeman,
  Munari, Campbell, Williams, Seabroke, Wyse, Parker, Bienaym{\'{e}}, Roeser,
  Gibson, Gilmore, Grebel, Helmi, Navarro, Burton, Cass, Dawe, Fiegert,
  Hartley, Russell, Saunders, Enke, Bailin, Binney, Bland-Hawthorn, Boeche,
  Dehnen, Eisenstein, Evans, Fiorucci, Fulbright, Gerhard, Jauregi, Kelz,
  Mijovi{\'{c}}, Minchev, Parmentier, Pe{\~{n}}arrubia, Quillen, Read, Ruchti,
  Scholz, Siviero, Smith, Sordo, Veltz, Vidrih, von Berlepsch, Boyle, \&
  Schilbach}]{Steinmetz2006}
Steinmetz, M., Zwitter, T., Siebert, A., {et~al.} 2006, AJ, 132, 1645

\bibitem[{Vera-Ciro {et~al.}(2014)Vera-Ciro, D'Onghia, Navarro, \&
  Abadi}]{VeraCiro2014}
Vera-Ciro, C., D'Onghia, E., Navarro, J., \& Abadi, M. 2014, ApJ, 794, 173

\bibitem[{Widrow {et~al.}(2012)Widrow, Gardner, Yanny, Dodelson, \&
  Chen}]{Widrow2012}
Widrow, L.~M., Gardner, S., Yanny, B., Dodelson, S., \& Chen, H.-Y. 2012, ApJL,
  750

\bibitem[{Williams {et~al.}(2013)Williams, Steinmetz, Binney, Siebert, Enke,
  Famaey, Minchev, de~Jong, Boeche, Freeman, Bienayme, Bland-Hawthorn, Gibson,
  Gilmore, Grebel, Helmi, Kordopatis, Munari, Navarro, Parker, Reid, Seabroke,
  Sharma, Siviero, Watson, Wyse, \& Zwitter}]{Williams2013}
Williams, M. E.~K., Steinmetz, M., Binney, J., {et~al.} 2013, MNRAS, 436, 101

\bibitem[{Yanny {et~al.}(2009)Yanny, Rockosi, Newberg, Knapp, Adelman-McCarthy,
  Alcorn, Allam, Prieto, Anderson, Anderson, Bailer-Jones, Bastian, Beers,
  Bell, Belokurov, Bizyaev, Blythe, Boroski, Brinchmann, Brinkmann, Brewington,
  Carey, Cudworth, Evans, Evans, Gates, G{\"{a}}nsicke, Gillespie, Gomez-Moran,
  Grebel, Greenwell, Jordan, Jordan, Harding, Harris, Hendry, Holder, Ivans,
  Ivezic, Jester, Johnson, Kent, Kleinman, Kniazev, Krzesinski, Kron,
  Kuropatkin, Lebedeva, Lee, Leger, Lepine, Levine, Lin, Long, Loomis,
  Malanushenko, Malanushenko, Margon, Martinez-Delgado, McGehee, Monet,
  Morrison, Munn, Neilsen, Nitta, Norris, Oravetz, Owen, Padmanabhan, Pan,
  Peterson, Pier, Platson, Fiorentin, Richards, Schlegel, Schneider, Schreiber,
  Schwope, Sibley, Simmons, Snedden, Smith, Stark, Stauffer, Steinmetz,
  Stoughton, SubbaRao, Szalay, Szkody, Thakar, Thirupathi, Tucker, Uomoto,
  Berk, Vidrih, Wadadekar, Watters, Wilhelm, Wyse, Yarger, \&
  Zucker}]{Yanny2009}
Yanny, B., Rockosi, C., Newberg, H., {et~al.} 2009, AJ, 137, 4377

\end{thebibliography}

\end{document}